\documentclass{article}
\usepackage[T1]{fontenc}

\usepackage{microtype}
\usepackage{graphicx}
\graphicspath{{../data/figures/}{figures/}{./}}
\usepackage{booktabs}
\usepackage{amsmath, amssymb, amsthm}

\usepackage{algorithm}
\usepackage{algorithmic}

\usepackage{float}
\usepackage{placeins}   
\usepackage{stfloats}
\usepackage{adjustbox}
\usepackage{array}
\usepackage{balance}    

\usepackage[accepted]{mlsys2025}

\theoremstyle{plain}

\theoremstyle{remark}
\newtheorem{remark}{Remark}

\newcommand{\kLane}{k_{\text{lane}}}
\newcommand{\kTotal}{k_{\text{total}}}
\newcommand{\Kpool}{K_{\text{pool}}}
\newcommand{\Sr}{S_r}
\newcommand{\Scup}{S_{\cup}}
\newcommand{\rhozero}{\rho_{0}}
\newcommand{\efsearch}{\texttt{efSearch}}
\newcommand{\nprobe}{\texttt{nprobe}}
\newcommand{\eqcost}{Equal cost and deadline.}
\newcommand{\alphapart}{\mbox{$\alpha$-partitioning}}
\newcommand{\holdingconst}[1]{holding #1 constant}

\makeatletter
\def\fps@figure{!tbp}
\def\fps@table{!tbp}
\makeatother

\makeatletter
\g@addto@macro\normalsize{%
  \setlength\abovecaptionskip{4pt}%
  \setlength\belowcaptionskip{2pt}%
}
\makeatother


\newcolumntype{L}[1]{>{\raggedright\arraybackslash}p{#1}}
\newcolumntype{C}[1]{>{\centering\arraybackslash}p{#1}}
\newcolumntype{R}[1]{>{\raggedleft\arraybackslash}p{#1}}

\usepackage{hyperref}
\hypersetup{
  pdftitle={Coordination-Free Lane Partitioning for Convergent ANN Search},
  pdfauthor={Carl Kugblenu and Petri Vuorimaa},
  pdfsubject={Vector Search, Approximate Nearest Neighbors, Distributed Systems},
  pdfkeywords={Vector Search, Approximate Nearest Neighbors, Distributed Systems, Parallel Search, Hedging}
}
\usepackage{xurl}      
\mlsystitlerunning{Coordination-Free Lane Partitioning for Convergent ANN Search}

\begin{document}
\flushbottom

\twocolumn[
\mlsystitle{Coordination-Free Lane Partitioning for Convergent ANN Search}

\mlsyssetsymbol{equal}{*}

\begin{mlsysauthorlist}
\mlsysauthor{Carl Kugblenu}{equal,aalto}
\mlsysauthor{Petri Vuorimaa}{equal,aalto}
\end{mlsysauthorlist}

\mlsysaffiliation{aalto}{Department of Computer Science, Aalto University, Espoo, Finland}

\mlsyscorrespondingauthor{Carl Kugblenu}{carl.kugblenu@aalto.fi}

\mlsyskeywords{Vector Search, Approximate Nearest Neighbors, Distributed Systems, Parallel Search, Hedging}

\vspace{.5\baselineskip}

\begin{abstract}
Production vector search systems often fan out each query across parallel lanes (threads, replicas, or shards) to meet latency service-level objectives (SLOs). In practice, these lanes rediscover the same candidates, so extra compute does not increase coverage. We present a coordination-free lane partitioner that turns duplication into complementary work at the same cost and deadline. For each query we (1) build a deterministic candidate pool sized to the total top-$k$ budget, (2) apply a per-query pseudorandom permutation, and (3) assign each lane a disjoint slice of positions. Lanes then return different results by construction, with no runtime coordination.

At equal cost with four lanes (total candidate budget 64), on SIFT1M (1M SIFT feature vectors) with Hierarchical Navigable Small World graphs (HNSW) recall@10 rises from 0.249 to 0.999 while lane overlap falls from nearly 100\% to 0\%. On MS MARCO (Microsoft MAchine Reading COmprehension; 8.8M passages) with HNSW, hit@10 improves from 0.200 to 0.601 and Mean Reciprocal Rank at 10 (MRR@10) from 0.133 to 0.330. For inverted file (IVF) indexes we see smaller but consistent gains (for example, +11\% on MS MARCO) by de-duplicating list routing. A microbenchmark shows planner overhead of \(\sim\)37~\textmu s per query (mean at the main setting) with linear growth in the number of merged candidates.

These results yield a simple operational guideline: size the per-query pool to the total budget, deterministically partition positions across lanes, and turn redundant fan-out into complementary coverage without changing budget or deadline.
\end{abstract}
]

\printAffiliationsAndNotice{\mlsysEqualContribution}

\section{Introduction}\label{sec:intro}

Production vector search systems rely on parallel processing to meet stringent latency and throughput requirements~\cite{dean2013tail}. In deployed engines, a coordinator scatters each query to shards/replicas and gathers partial results to meet tail SLOs~\citep{dean2013tail,wang2021milvus}; randomized load balancing via the power of two choices offers a simple baseline for work distribution~\cite{mitzenmacher2002power}. If properly utilized, parallel lanes should yield distinct candidates. Instead, we observe the opposite: on SIFT1M, four HNSW lanes (per-lane 16; total 64) achieve recall@10 of only 0.249 versus a single-index search with the same total budget (64). On MS MARCO (8.8M documents), the same configuration yields hit@10 of 0.200 compared to 0.601 for single-index. The lanes expend identical compute but deliver near-identical results.

Parallelism is necessary but not sufficient. Classic randomized load balancing results show small, local choices can greatly improve distribution of work~\cite{mitzenmacher2002power}, while system-level hedging helps with tails but often returns duplicates under contention~\cite{dean2013tail,primorac2021hedge}. Orthogonal lines of work improve the underlying index or hardware: graph partitioning for nearest neighbor search~\cite{gottesburen2025unleashing}, SSD-aware scan/seek patterns~\cite{shim2025turbocharging}, and disaggregated memory for RAG pipelines~\cite{liu2025efficient}. Our planner targets a different lever: it reallocates the \emph{budget} across lanes so the same compute explores complementary regions. This contrasts with index-partitioning methods, which change the layout of the data structure itself, while we only change how a single query spends its budget.

The root cause is algorithmic convergence. HNSW's greedy beam search from a shared entry point forces all lanes onto the same traversal path~\cite{malkov2018efficient}. Even with different starting nodes, the hierarchical structure funnels searches toward the same neighborhoods. Munyampirwa et al. formalize hub-driven convergence \cite{munyampirwa2024down}; we show its deployment consequence: parallel lanes duplicate work, which our planner converts into disjoint coverage under the same budget.

IVF: At query time the system selects the closest coarse centroids and scans only those inverted lists~\citep{jegou2010product}. Under multi-lane fan-out, independently launched probes tend to pick the same high-scoring centroids, so list-level overlap is high even if document-level diversity exists. We define a convergence coefficient as the Jaccard overlap at baseline parallelism. On both SIFT1M and MS MARCO, across HNSW and IVF, we measure this coefficient at near 1.00, indicating near-perfect duplication. Comprehensive evaluations across graph/vector methods reinforce how structural choices drive overlap and headroom for improvement~\cite{azizi2025graph}. Our results align with this picture: IVF shows higher intra-list diversity than HNSW but still leaves recoverable list-level duplication.

This waste is a systems-level artifact, not an algorithmic flaw. The approximate nearest neighbor (ANN) algorithms are correct; the deployment strategy is naive. We address this with \alphapart{}: a coordination-free planner that makes lane work disjoint by construction. The method builds one deterministic candidate pool per query, applies a pseudorandom function (PRF) to break correlation, then assigns each lane dedicated positions for its budget, backfilling from pool order for efficiency. The mechanism gives:

\begin{itemize}
\item Disjoint coverage by construction: With a pool equal to the total budget and full dedication, lane selections are disjoint by construction (Remark~\ref{rem:disjoint}).
\item Equal-cost discipline: Same total budget, same deadline, same traversal work.
\item Coordination-free operation: Lanes share only PRF definition and query seed; no runtime inter-lane messages.
\item Operational simplicity: One metric predicts gains; one rule for sizing.
\end{itemize}

\paragraph{Contributions:}
\begin{enumerate}
\item We diagnose convergent traversal via a Jaccard coefficient (\ref{sec:rho0}) and quantify waste in production parallelism, validating against standard deployment patterns (\ref{sec:validation}).
\item We introduce \alphapart{}: a coordination-free mechanism that converts duplication into complementary coverage at equal cost (\ref{sec:method}).
\item We prove disjointness under full dedication (Remark~\ref{rem:disjoint}) and provide a coverage model validated across 2 algorithms $\times$ 2 datasets $\times$ 2 scales (1M, 8.8M).
\item We provide an operational playbook: measure overlap, set dedication, size the pool (\ref{sec:deployment}).
\end{enumerate}

\noindent\fbox{%
\parbox{\linewidth}{%
\textbf{Equal-cost, equal-deadline invariant.}
All comparisons in this paper fix: (i) the total candidate budget (sum across lanes), (ii) the deadline/SLO, and (iii) traversal work equal to a single-index baseline. Only the partition knob changes how the per-query budget is split across lanes. Unlike classic hedging (which often duplicates work and then discards late results), our planner assigns complementary work up front so late arrivals still add unique coverage.
}%
}

\vspace{0.5\baselineskip}

\noindent\textit{Results preview:} Equal-cost gains across two datasets and two index families, with zero overlap at full dedication and parity to the single-index ceiling.

\section{Problem Setup and Notation}
\label{sec:setup}

We use a small set of symbols for the remainder of the paper; Table~\ref{tab:notation} summarizes them, and we stick to this notation consistently.

\begin{table}[t]
\caption{Notation used throughout the paper.}
\label{tab:notation}
\centering
\begin{tabular}{l l}
\toprule
Symbol & Meaning \\
\midrule
$M$ & Number of parallel lanes \\
$k_{\text{lane}}$ & Per-lane candidate budget \\
$k_{\text{total}} = M\cdot k_{\text{lane}}$ & Total candidate budget \\
$K_{\text{pool}}$ & Size of the per-query candidate pool \\
$S_r$ & Set returned by lane $r$ (size $k_{\text{lane}}$) \\
$S_{\cup}$ & $\bigcup_{r=1}^{M} S_r$ (union of lane results) \\
$\rho_0$ & Baseline convergence coefficient (Jaccard) \\
$\alpha$ & Dedicated fraction for partitioning \\
\texttt{efSearch} & HNSW beam size at query time \\
\texttt{nprobe} & IVF number of coarse lists probed \\
\bottomrule
\end{tabular}
\end{table}

\noindent\fbox{%
\parbox{\linewidth}{%
\textbf{Equal-cost, equal-deadline (symbolic).}
All comparisons fix: (i) $\kTotal = M\cdot \kLane$, (ii) the deadline/SLO, and (iii) traversal work parity with the single-index baseline. Only $\alpha$ changes how the per-query budget is split across lanes.
}%
}

\subsection{Multi-Lane Protocol}

Production deployments use parallel lanes in multiple contexts:
\begin{itemize}
\item Hedging: Primary/secondary replicas to mitigate tail latency~\cite{dean2013tail,primorac2021hedge}
\item Intra-node parallelism: Multi-threaded HNSW traversals or IVF probes
\item Geographic distribution: Multi-region active-active serving
\item Sharding: Index partitions with fan-out queries
\end{itemize}

The standard protocol broadcasts a query to $M$ lanes, each returning $\kLane$ candidates:

\begin{algorithmic}
\FOR{$r \in \{0, \ldots, M-1\}$}
    \STATE $\Sr \gets \text{lane}[r].\text{search}(q, \kLane)$
\ENDFOR
\STATE \textbf{return} $\text{top}_k(\text{merge}(\text{results}), k)$
\end{algorithmic}

Ideally, this yields $M \times \kLane$ distinct candidates. In practice, lanes converge.

\subsection{Convergence Coefficient}
\label{sec:rho0}

\paragraph{Convergence coefficient $\rhozero$.}
Let $\Sr$ be the size-$\kLane$ set returned by lane $r$ under baseline parallelism ($\alpha{=}0$). We define the baseline convergence coefficient $\rhozero$ as
\[
\rhozero \;=\; \frac{\bigl|\bigcap_{r=1}^{M} \Sr\bigr|}{\bigl|\bigcup_{r=1}^{M} \Sr\bigr|}.
\]
High $\rhozero$ indicates that parallel budget is being duplicated rather than expanding distinct coverage. 
\emph{Toy example} ($M{=}3$): if
$S_1{=}\{a,b,c\}$, $S_2{=}\{a,b,d\}$, $S_3{=}\{a,b,e\}$, then 
$\lvert\cap \Sr\rvert{=}2$, $\lvert\cup \Sr\rvert{=}5$, so $\rhozero{=}2/5{=}0.4$.

On SIFT1M and MS MARCO, we measure $\rhozero \approx 1.00$ for both HNSW and IVF, confirming that current parallel strategies are fundamentally inefficient.

\subsection{Problem validation in production}
\label{sec:validation}

Scatter-gather parallelism is how real vector stores meet SLOs: a coordinating node fans out a query to shards/replicas and merges partial results. In Elasticsearch, the coordinating node requests work from each shard and reduces shard responses before returning a single result; for kNN search each shard returns candidates and the coordinator merges to the final top-$k$~\citep{elastic-knn-numcands}. Milvus implements the same pattern explicitly: QueryNodes reduce across segments, StreamingNodes reduce across QueryNodes, and the Proxy performs the final merge~\citep{wang2021milvus}.

For IVF, the standard procedure is to select the \nprobe{} nearest coarse centroids and then scan those inverted lists~\citep{jegou2010product,johnson2019billion}. The Faiss library provides functionality to search caller-specified list sets~\citep{douze2024faiss}, which we use to route each lane's disjoint slice without modifying the index.

Together with recent evidence that HNSW trajectories converge to graph hubs~\citep{munyampirwa2024down}, these architectures explain why naive parallelism re-discovers the same candidates: independent lanes route to the \emph{same} shards/lists and traverse \emph{similar} neighborhoods. Our $\rho_0 \!\approx\! 1.00$ measurements quantify this duplication under standard production configurations.

\section{Method: \texorpdfstring{\alphapart{}}{alpha-partitioning}}
\label{sec:method}

\subsection{Pool-Then-Partition Recipe}

\begin{figure*}[!t]         
\centering
\includegraphics[width=\textwidth]{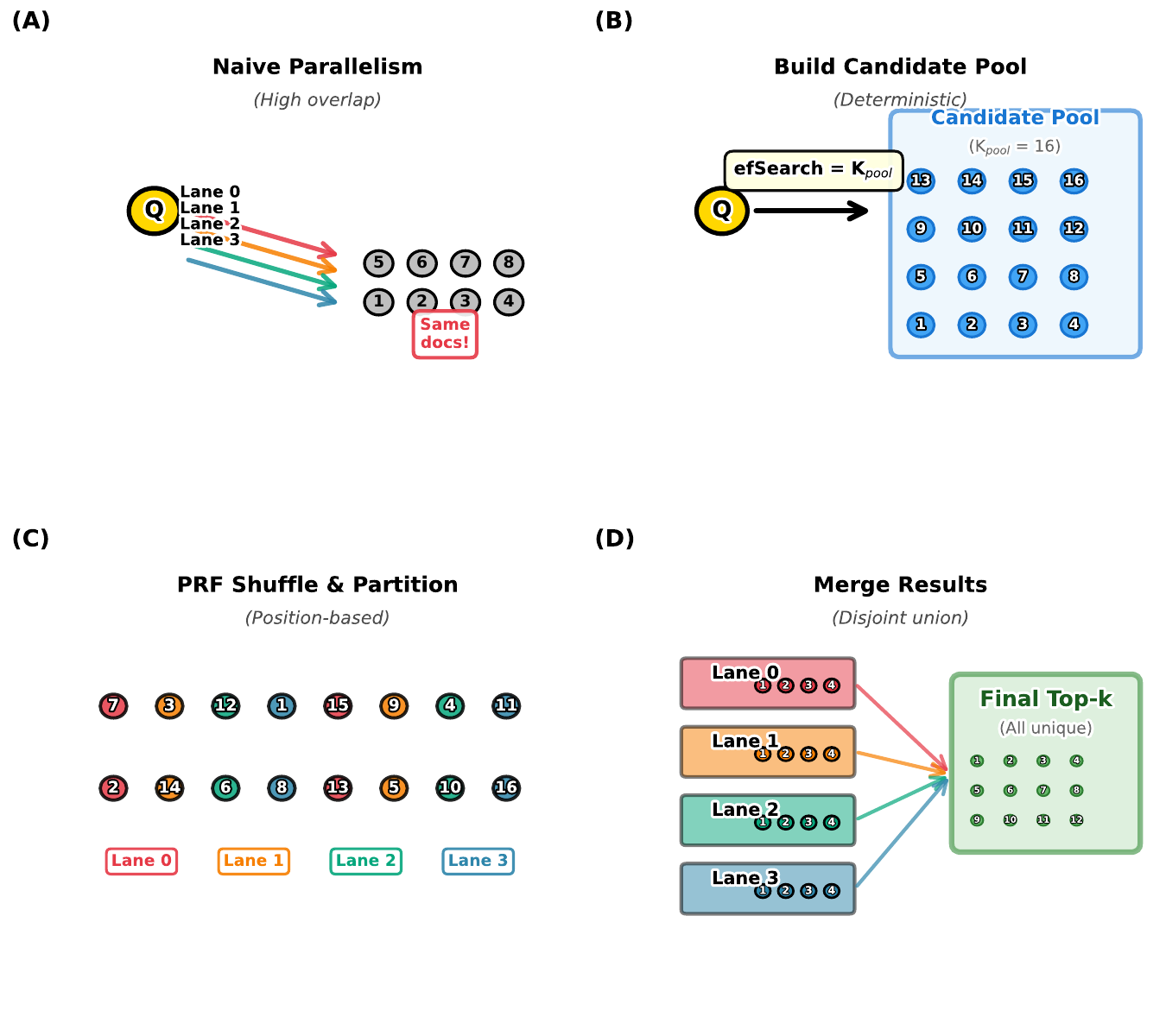}
\vspace{-0.5\baselineskip}  
\caption{\alphapart{} Workflow. (A) Naive parallelism: high overlap, wasted budget. (B) Build deterministic candidate pool. (C) PRF shuffle and position-based partition. (D) Disjoint union merge yields zero overlap.}
\label{fig:conceptual}
\end{figure*}

\alphapart{} transforms naive parallelism into coordination-free disjoint search via three steps (Figure~\ref{fig:conceptual}):

\begin{enumerate}
\item Build deterministic pool: Generate a candidate pool $C_q \subset U$ of size $\Kpool$ using the base ANN algorithm (e.g., set \efsearch=$\Kpool$ in HNSW or \nprobe\ to cover $\Kpool$ candidates in IVF).

\item PRF shuffle: Apply a pseudorandom function $\text{PRF}(q, \text{docid})$ to sort $C_q$ into a deterministic permutation $\pi_q[0], \ldots, \pi_q[\Kpool-1]$~\cite{nist_prf}; the per-query keyed hashing is analogous in spirit to consistent hashing for stable partitions~\cite{karger1997consistent}. We use a 64-bit multiplicative hash (splitmix64-based) keyed by query ID that completes in sub-microsecond per document. This PRF is deterministic per query and requires no cross-lane communication.

\item Position-based partition: Define dedicated quota $k_{\text{ded}} = \lfloor \alpha \cdot \kLane \rfloor$ and shared quota $k_{\text{shr}} = \kLane - k_{\text{ded}}$. Lane $r$ takes positions $\{r, r+M, r+2M, \ldots, r+(k_{\text{ded}}-1)M\}$ (dedicated) and backfills from positions $[k_{\text{ded}} \cdot M, k_{\text{ded}} \cdot M + k_{\text{shr}} - 1]$ (shared). This construction implies the zero-overlap property proven in Remark~\ref{rem:disjoint}.
\end{enumerate}

\paragraph{Reference implementation:}

\begin{algorithmic}
\STATE \textbf{function} \textsc{AlphaPartition}$(C_{\text{pool}}, M, \kLane, \alpha, r)$
\STATE \quad $k_{\text{ded}} \gets \lfloor \alpha \cdot \kLane \rfloor$
\STATE \quad $\text{dedicated} \gets C_{\text{pool}}[r::M][:k_{\text{ded}}]$ \COMMENT{congruence class $r \mod M$}
\STATE \quad $\text{chosen} \gets \text{list}(\text{dedicated})$
\FOR{$c$ in $C_{\text{pool}}$}
    \IF{$|\text{chosen}| \ge \kLane$}
        \STATE \textbf{break}
    \ENDIF
    \IF{$c \notin \text{chosen}$}
        \STATE $\text{chosen}.\text{append}(c)$
    \ENDIF
\ENDFOR
\STATE \textbf{return} $\text{chosen}[:\kLane]$
\end{algorithmic}

\subsection{Algorithm Integration}

HNSW: Two-phase approach. (1) Enumerate pool: run HNSW search with \efsearch=$\Kpool$ to generate $C_q$. (2) Partition: apply PRF, assign positions, each lane rescores its subset. Work comparison: $W_{\text{pool}}$ (enumerate $\Kpool$ nodes) + $O(\kTotal)$ (PRF/partition/merge) $\approx W_{\text{single}}$ (\efsearch=$\kTotal$). 

IVF: Partition coarse list IDs. At query time, select $\Kpool$ candidates across multiple lists, PRF-order them, partition by position. We use FAISS functionality to pass each lane its pre-partitioned list IDs, preserving identical per-list scan work \holdingconst{\nprobe{}}~\citep{douze2024faiss}. Merge is $O(\kTotal \log M)$.

Coordination-free: Lanes share only PRF definition (a static function) and query seed. No runtime messages. Each lane independently computes its positions and selections. Runtime coordination is not required; practical straggler policies are discussed in \ref{sec:stragglers}.

\section{Theory}

\subsection{A one-line guarantee at full dedication}

\begin{remark}[Disjointness at full dedication]
\label{rem:disjoint}
Let $\Kpool \ge \kTotal = M \cdot \kLane$ and $\alpha{=}1$. After a per-query PRF permutation of the pool, assign lane $r$ the positions $\{\,r,\, r{+}M,\, \dots,\, r{+}(\kLane{-}1)M\,\}$. These position sets are disjoint congruence classes modulo $M$ within the first $\kTotal$ items, so lane outputs are pairwise disjoint and $\lvert \Scup\rvert = \kTotal$ by construction.
\end{remark}

\subsection{Coverage accounting (just counting)}

Under feasibility ($\Kpool \ge M k_{\text{ded}} + k_{\text{shr}}$) with $k_{\text{ded}}=\lfloor \alpha \kLane \rfloor$ and $k_{\text{shr}}=\kLane-k_{\text{ded}}$,
\begin{equation}
\label{eq:coverage}
|\Scup(\alpha)| = M k_{\text{ded}} + k_{\text{shr}} = \kLane\bigl(1 + \alpha(M-1)\bigr).
\end{equation}
Intuition: $\alpha$ converts a fraction of each lane's budget into dedicated, non-overlapping positions; the rest is a shared suffix.

\subsection{Operational predictor from \texorpdfstring{$\rho_0$}{rho0}}

Let $U_0 = |\Scup(0)|$ be distinct coverage at $\alpha=0$, and define the baseline convergence coefficient
\[
\rhozero \;=\; \frac{\bigl|\bigcap_{r=1}^{M} \Sr\bigr|}{\bigl|\bigcup_{r=1}^{M} \Sr\bigr|}.
\]
Approximating $U_0 \approx \kLane \bigl[1 + (M-1)(1-\rhozero)\bigr]$, the lift from $\alpha{=}0$ to $\alpha{=}1$ at equal cost is
\begin{equation}
\label{eq:op-predictor}
\text{Gain} \;\approx\; \frac{\kTotal}{U0} \;\approx\; \frac{M}{1 + (M-1)(1-\rhozero)}.
\end{equation}
Checks: if $\rhozero \to 1$, then $\text{Gain} \to M$; if $\rhozero=0$, then $\text{Gain}=1$.

\subsection{Sizing rule}

Set $\Kpool = \kTotal$ to realize the zero-overlap union in Remark~\ref{rem:disjoint}. Under-pooling ($\Kpool < \kTotal$) degrades recall proportionally, which we validate empirically.

\section{Experimental Setup}

\subsection{Datasets and Embeddings}

SIFT1M: 1,000,000 vectors, 128d, L2 distance. Ground truth computed via Faiss \texttt{IndexFlatL2}. Standard ANN benchmark~\cite{johnson2019billion}.

MS MARCO Passages: 8,841,823 documents. Embeddings: BAAI/bge-small-en-v1.5 (384d), unit-normalized. Cosine similarity (L2-equivalent on unit vectors) for HNSW, inner product for IVF. Dev queries (6,980) with official qrels.

\subsection{Algorithms}

HNSW: M=32, efConstruction=200. During pooling, \efsearch=$\Kpool$. Implementation via hnswlib.

IVF-Flat: nlist=4096, trained on 160K samples per seed. Each lane uses identical per-list scan work via Faiss search functionality.

\subsection{Protocol}

Main comparison: M=4, $\kLane$=16 $\Rightarrow \kTotal$=64. All plots and tables obey the equal-cost, equal-deadline invariant (see box in \ref{sec:intro}). Equal-cost invariant: same total budget, same deadline, only $\alpha$ changes partition strategy.

$\alpha$ sweep: $\{0, 0.25, 0.5, 0.75, 1.0\}$. Seeds: $\{42, 123, 789\}$ for reproducibility.

$\Kpool$: $\kTotal$ unless noted. Pool-size ablations use $\{0.8, 0.9, 1.0, 1.1, 1.25, 1.5\} \times \kTotal$.

Reporting: mean $\pm$ std.\footnote{Sampling variance decreases as total covered candidates grow; in practice we observe tight intervals with $k_{\text{total}}$ at our settings.}

Our setup is CPU-centric; SSD-aware vector database designs can further change scan/seek trade-offs~\cite{shim2025turbocharging}, which we view as complementary to the budget-partitioning we study.

\subsection{Metrics}

SIFT1M: recall@10, overlap $\rho$ (Jaccard).

MS MARCO HNSW: hit@10 (fraction of queries with $\ge$1 relevant doc in top-10), MRR@10 (Mean Reciprocal Rank at 10), overlap $\rho$.

MS MARCO IVF: recall@10 (fraction of relevant docs retrieved), overlap $\rho$.

\noindent\textit{Reproducibility note.} Equal-cost instrumentation (node visits, list scans, and planner timing) and the exact command lines are in the Appendix document.

\section{Results}\label{sec:results}

All comparisons obey the equal-cost, equal-deadline contract (see box in \ref{sec:intro}).

\subsection{SIFT1M HNSW: Extreme Convergence to Large Gains}

\begin{figure}[t]
\centering
\includegraphics[width=\columnwidth]{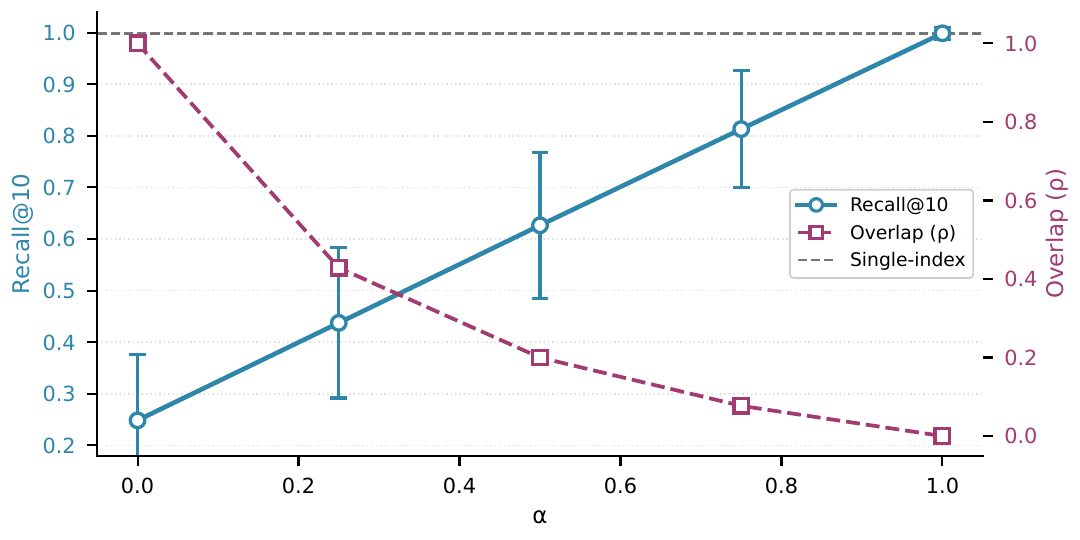}
\caption{SIFT1M, HNSW, $\alpha$-sweep with $M{=}4$, $\kLane{=}16$ ($\kTotal{=}64$). The gray dashed line marks the single-index ceiling at equal cost (in our run this equals the $\alpha{=}1$ mean, \emph{0.999}); $\alpha$ increases coverage monotonically from 0.249 to 0.999 while overlap drops from 1.00 to 0.00. \eqcost}
\label{fig:sift_hnsw_sweep}
\end{figure}

Figure~\ref{fig:sift_hnsw_sweep} shows the fundamental trade-off. As $\alpha$ increases from 0.0 to 1.0, overlap drops from 1.00 to 0.00 while recall climbs from 0.249 to 0.999.

\begin{table}[t]
\centering
\small
\setlength{\tabcolsep}{4pt}
\begin{adjustbox}{max width=\columnwidth}
\begin{tabular}{lccccc}
\toprule
Config & $\rhozero$ & recall@10 $\alpha$=0 & recall@10 $\alpha$=1 & Single & Gain \\
\midrule
M=4, $\kLane$=16 & 1.00$\pm$0.00 & 0.249$\pm$0.001 & 0.999$\pm$0.001 & 0.999$\pm$0.001 & +302\% \\
\bottomrule
\end{tabular}
\end{adjustbox}
\caption{SIFT1M HNSW Main Results (M=4, $\kLane$=16, $\kTotal$=64)}
\label{tab:sift_hnsw_main}
\end{table}

Table~\ref{tab:sift_hnsw_main} quantifies the impact. The +302\% gain is the difference between a system that fails to scale and one that succeeds.

\subsection{SIFT1M IVF: High Baseline to Modest Headroom}

\paragraph{IVF: Lower document-level $\rhozero$, but recoverable list-level duplication.}
As discussed in \ref{sec:intro}, IVF shows higher intra-list diversity than HNSW. However, recoverable duplication lives at the \emph{routing boundary}. \alphapart{} assigns each lane disjoint positions in a PRF-shuffled ordering of coarse lists, so lanes scan different lists first instead of piling into the same ones. This converts redundant list scans into complementary coverage \holdingconst{\nprobe{}} and with identical per-list scan work. Empirically, this yields modest but reliable gains at equal cost (e.g., +16\% on SIFT1M-IVF and +11\% on MS MARCO-IVF), smaller than HNSW because document-level $\rhozero$ is already lower, but still ``free quality'' recovered from list-level overlap.

\noindent\textbf{General principle validated beyond HNSW.}
In HNSW, correlation presents as convergent \emph{traversal} to the same nodes; in IVF, as convergent \emph{routing} to the same lists. In both cases, budget partitioning over a deterministically ordered pool enforces disjoint coverage at equal cost. Seeing consistent gains on IVF confirms the planner is not tied to a single index family.

Table~\ref{tab:sift_ivf_main} shows $\alpha=0$ already achieves 0.838 recall, indicating IVF has intra-list diversity. $\alpha=1$ lifts this to 0.971 (+16\%), a quality gain at zero additional cost. See \ref{sec:msmarco_ivf} for why IVF's lower document-level $\rhozero$ still leaves recoverable list-level duplication.

\begin{table}[t]
\centering
\small
\setlength{\tabcolsep}{4pt}
\begin{adjustbox}{max width=\columnwidth}
\begin{tabular}{lcccc}
\toprule
Metric & $\rhozero$ & recall@10 $\alpha$=0 & recall@10 $\alpha$=1 & Gain \\
\midrule
IVF-Flat & 1.00$\pm$0.00 & 0.838$\pm$0.002 & 0.971$\pm$0.001 & +16\% \\
\bottomrule
\end{tabular}
\end{adjustbox}
\caption{SIFT1M IVF Results. High baseline (0.838 at $\alpha=0$) indicates intra-list diversity; \alphapart{} recovers list-level overlap as complementary coverage.}
\label{tab:sift_ivf_main}
\end{table}

\subsection{MS MARCO HNSW: Equal-Cost Parity at Scale}

\begin{figure}[t]
\centering
\includegraphics[width=\columnwidth]{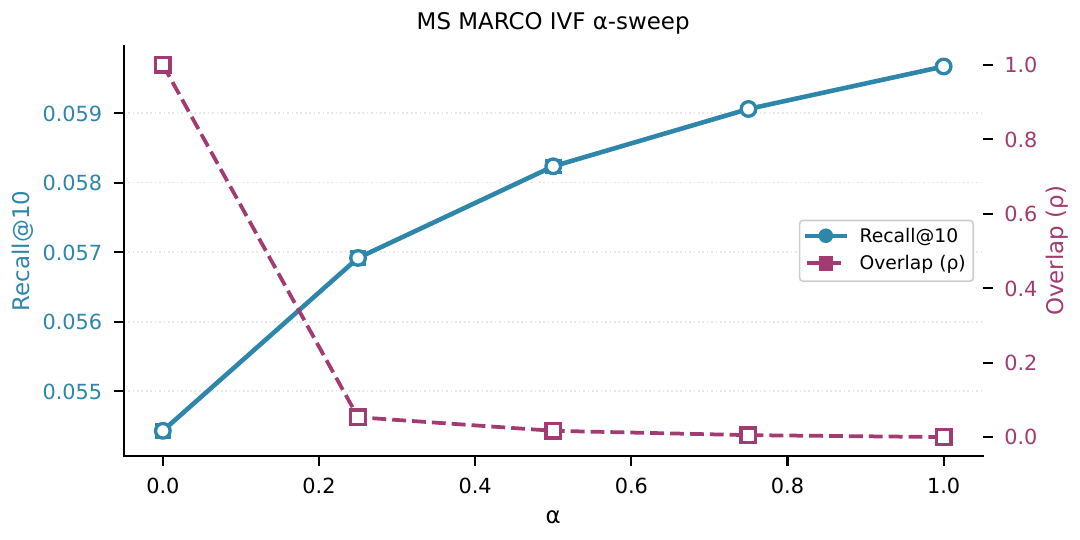}
\caption{MS MARCO (8.8M), IVF-Flat, $\alpha$-sweep. Setup: $M{=}4$, $\kLane{=}16$, $\kTotal{=}64$. Left axis: recall@10; right axis: overlap. Despite lower document-level $\rhozero$ (intra-list diversity), \alphapart{} recovers list-level duplication for a +11\% gain at equal cost (same total candidate budget and deadline).}
\label{fig:msmarco_ivf_alpha}
\end{figure}

\begin{figure}[t]
\centering
\includegraphics[width=\columnwidth]{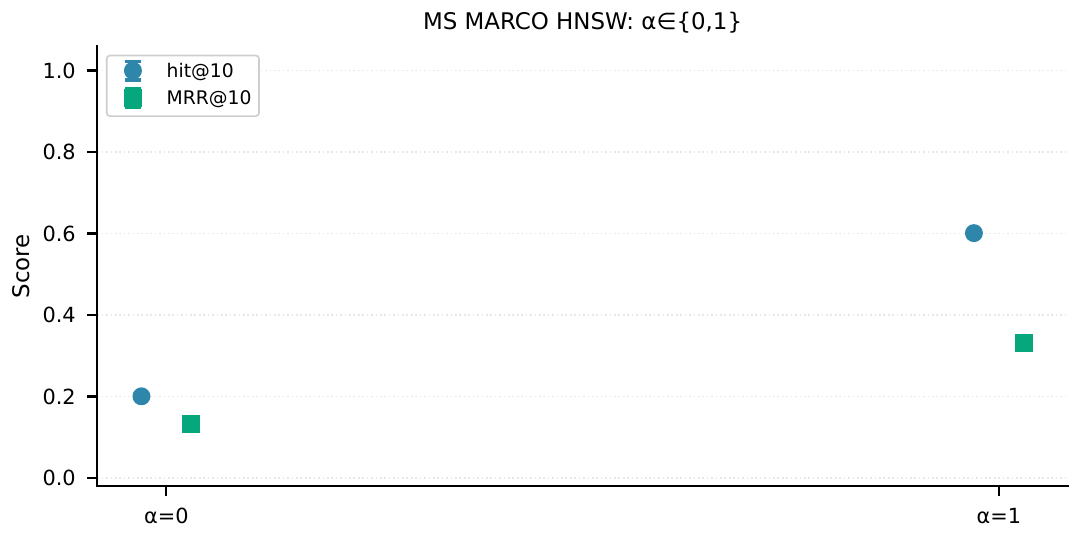}
\caption{MS MARCO (8.8M), HNSW at $\alpha\in\{0,1\}$ with $M{=}4$, $\kLane{=}16$. Metrics: hit@10 and MRR@10 (mean $\pm$ std). $\alpha{=}1$ reaches the single-index ceiling (+200\% hit@10, +148\% MRR@10) at the same budget and deadline.}
\label{fig:msmarco_hnsw_alpha01}
\end{figure}

Figures~\ref{fig:msmarco_ivf_alpha} and \ref{fig:msmarco_hnsw_alpha01}, together with Table~\ref{tab:msmarco_hnsw}, demonstrate production-scale validation on 8.8M documents with BGE embeddings.

\begin{table}[t]
\centering
\small
\setlength{\tabcolsep}{4pt}
\begin{adjustbox}{max width=\columnwidth}
\begin{tabular}{lccccc}
\toprule
Metric & $\rhozero$ & $\alpha$=0 & $\alpha$=1 & Single & Gain \\
\midrule
hit@10 & 1.00 & 0.200 & 0.601 & 0.601 & +200\% \\
MRR@10 & 1.00 & 0.133 & 0.330 & 0.330 & +148\% \\
\bottomrule
\end{tabular}
\end{adjustbox}
\caption{MS MARCO HNSW (8.8M, M=4, $\kLane$=16, $\kTotal$=64)}
\label{tab:msmarco_hnsw}
\end{table}

Parity with single-index at equal cost demonstrates zero-waste recovery. The +200\% hit@10 gain confirms that the SIFT1M pattern holds at production scale.

\subsection{MS MARCO IVF: Moderate Gains at Zero Cost}
\label{sec:msmarco_ivf}

As discussed, IVF shows lower document-level ($\rhozero$) due to intra-list diversity; we still recover list-level duplication with \alphapart{}. Table~\ref{tab:msmarco_ivf} shows recall@10 $0.054 \to 0.060$ (+11\%). Consistency with SIFT1M IVF: moderate gains predicted by lower doc-level $\rhozero$.

\begin{table}[t]
\centering
\small
\setlength{\tabcolsep}{4pt}
\begin{adjustbox}{max width=\columnwidth}
\begin{tabular}{lcccc}
\toprule
Metric & $\rhozero$ & recall@10 $\alpha$=0 & recall@10 $\alpha$=1 & Gain \\
\midrule
IVF-Flat & 1.00 & 0.054 & 0.060 & +11\% \\
\bottomrule
\end{tabular}
\end{adjustbox}
\caption{MS MARCO IVF (8.8M)}
\label{tab:msmarco_ivf}
\end{table}

\subsection{Coverage Model Validation}

\begin{figure}[t]
\centering
\includegraphics[width=\columnwidth]{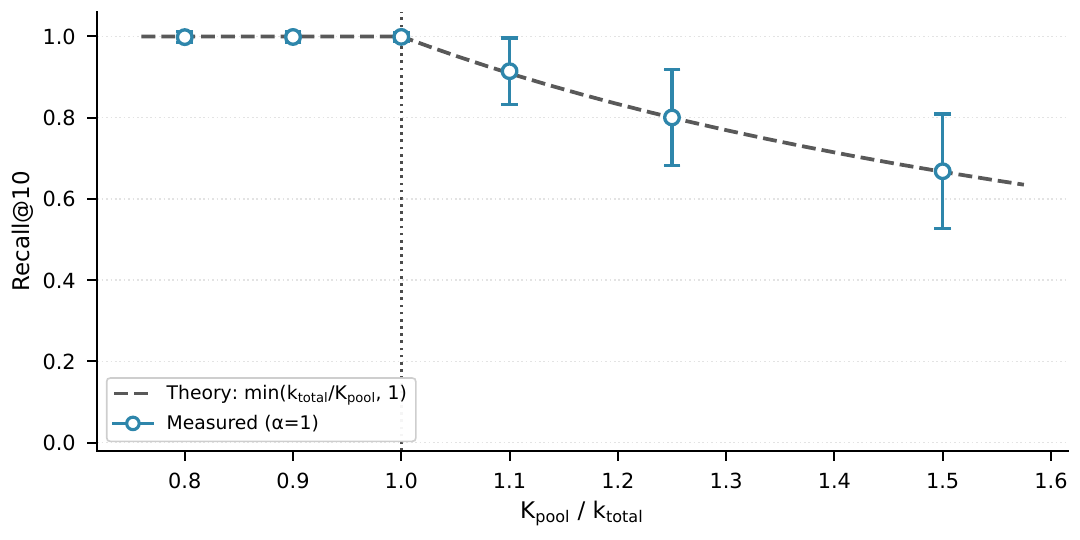}
\caption{Coverage validation on SIFT1M at $\alpha{=}1$. X: $\Kpool/\kTotal$; Y: recall@10. Measured points track the predicted $ \min(\kTotal/\Kpool,1)$ curve. $\Kpool{=}\kTotal$ maximizes quality with zero overlap; under-pooling degrades recall. Equal-cost evaluation.}
\label{fig:coverage_validation}
\end{figure}

Figure~\ref{fig:coverage_validation} validates the theoretical coverage relationship. Measured recall at $\alpha=1$ tracks $\kTotal / \Kpool$ prediction across pool sizes $\{0.8, 0.9, 1.0, 1.1, 1.25, 1.5\} \times \kTotal$.

Key finding: $\Kpool = \kTotal$ is optimal (maximizes recall with zero overlap). Under-pooling ($\Kpool < \kTotal$) degrades quality.

\subsection{Lane Scaling}

\begin{figure}[t]
\centering
\includegraphics[width=\columnwidth]{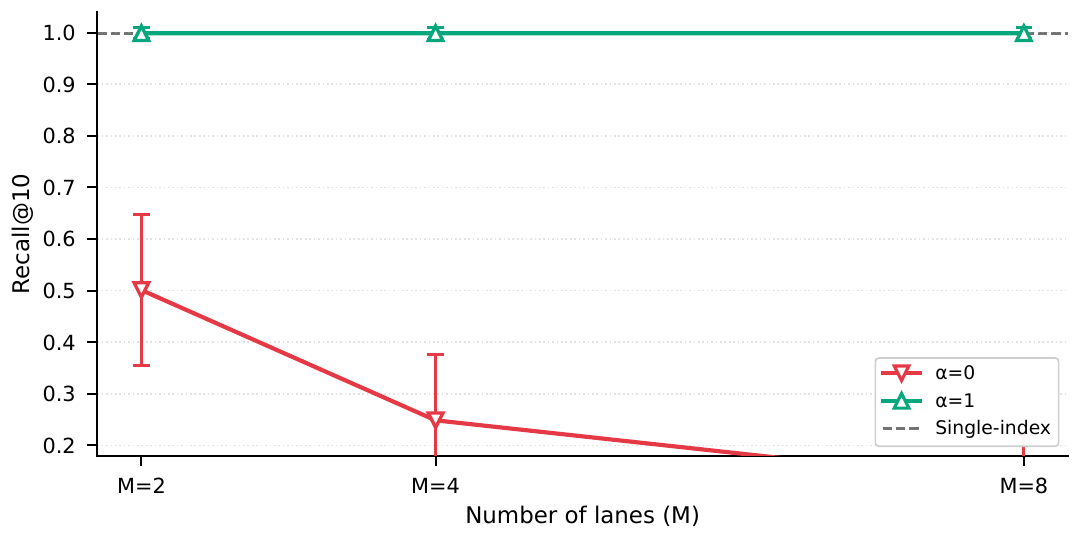}
\caption{Lane scaling on SIFT1M HNSW. $M\in\{2,4,8\}$, $\kLane{=}16$. Naive ($\alpha{=}0$) collapses as $M$ grows; $\alpha{=}1$ remains at the single-index ceiling with $\rho{=}0$. Equal total budget $\kTotal{=}M\cdot\kLane$ in all settings.}
\label{fig:lane_scaling}
\end{figure}

Figure~\ref{fig:lane_scaling} provides direct empirical validation of the "Tail at Scale" motivation~\cite{dean2013tail}.

$\alpha=0$ (naive): Recall plummets from 0.312 (M=2) to 0.084 (M=8). This is the "tail at scale" effect: theoretical benefit of more compute is offset by increasing cost of redundant work and system variability.

$\alpha=1$ (partitioned): Stable performance. Recall remains high and flat as M increases. This is visual proof that by eliminating computational redundancy, \alphapart{} creates a scalable, tail-tolerant system.

\begin{table}[t]
\centering
\small
\setlength{\tabcolsep}{4pt}
\begin{adjustbox}{max width=\columnwidth}
\begin{tabular}{lccc}
\toprule
M & $\alpha$=0 & $\alpha$=1 & Single \\
\midrule
2 & 0.312 & 0.668 & 0.668 \\
4 & 0.249 & 0.999 & 0.999 \\
8 & 0.084 & 0.665 & 0.668 \\
\bottomrule
\end{tabular}
\end{adjustbox}
\caption{Lane Scaling (SIFT1M, $\alpha \in \{0,1\}$). "Single" denotes the single-index result at the same total budget for each $M$ (e.g., $M{=}8 \Rightarrow k_{\text{total}}{=}128$).}
\label{tab:lane_scaling}
\end{table}

\subsection{Systems Summary}

\textbf{Latency.} Our comparisons hold total traversal work constant (equal cost and deadline), so end-to-end latency should remain flat; the planner adds only $O(k_{\text{total}})$ PRF/partition/merge work. In a microbenchmark at $M{=}4$, $k_{\text{lane}}{=}16$ ($k_{\text{total}}{=}64$) the planner cost is $36.8\,\mu\mathrm{s}$ mean ($p50{\approx}36.3\,\mu\mathrm{s}$, $p95{\approx}37.6\,\mu\mathrm{s}$), i.e., negligible relative to ANN traversal.

Equal-cost verified: Node-visit parity (HNSW), list-scan parity (IVF). Counters and exact scripts for node-visit parity (HNSW) and list-scan parity (IVF) are provided in the Appendix document.

A full latency profile and instrumentation scripts are provided in the accompanying artifact.

\FloatBarrier
\section{Related Work}
\label{sec:related}

Optimizations for approximate nearest neighbor search can be viewed at distinct layers of the system stack. Our work introduces a novel technique at the \emph{query execution layer}, complementing advances at the algorithm, data, and hardware layers.

\paragraph{The Algorithm Layer: Index Structures.}
HNSW~\cite{malkov2018efficient} and related graph methods like DiskANN~\cite{subramanya2019diskann} and SPANN~\cite{chen2021spann} focus on creating efficient index structures. Recent comprehensive studies systematically evaluate the trade-offs between these methods, highlighting regimes where each excels~\cite{azizi2025graph}. While these innovations make any single search traversal more efficient, they do not address how to coordinate multiple parallel traversals. Our \alphapart{} acts as a meta-algorithm that ensures the efficiency gains from these advanced indexes are not squandered on redundant work in production fan-out scenarios.

\paragraph{The Data Layer: Data Partitioning vs. Budget Partitioning.}
A common strategy for scaling search is to partition the \emph{data} via sharding or graph partitioning~\cite{gottesburen2025unleashing}. These methods split the index itself across nodes, which is essential for datasets that exceed single-machine memory. In contrast, we partition the \emph{per-query candidate budget}. This makes our approach a lightweight, state-free coordination mechanism that can be layered on top of any data partitioning scheme. When a query is fanned out to multiple data shards, our method can be applied locally on each shard to de-duplicate work among parallel threads, turning a local inefficiency into a global gain.

\paragraph{The Hardware Layer: System-Aware Optimization.}
Hardware-aware designs further optimize performance by tailoring algorithms to the underlying system, for instance via SSD-aware I/O patterns~\cite{shim2025turbocharging} or disaggregated memory for large models~\cite{liu2025efficient}. These are complementary to our work. An efficient I/O path makes each unit of traversal work cheaper, while our method ensures that fewer units of work are wasted. Combining these approaches yields multiplicative benefits: the system performs cheaper work, and less of it is redundant.

\paragraph{The Query Execution Layer: Redundancy and Robustness.}
Our work re-frames the role of redundancy in parallel execution. System-level hedging reduces tail latency by issuing duplicate requests, accepting computational overhead as the cost of robustness~\cite{dean2013tail,primorac2021hedge}. We demonstrate that this trade-off is often unnecessary. By deterministically partitioning the search space, \alphapart{} preserves the latency benefits of parallel execution; as the first-arriving lane can still serve the request while converting redundant compute into complementary coverage. This transforms parallelism from a mere robustness mechanism into a tool for improving quality at the same SLO. This principle connects to speculative execution in databases and operating systems~\cite{nightingale2005speculative,wei2024rome}, but our approach avoids duplicate work proactively rather than resolving conflicts post-hoc.

\paragraph{Relationship to IR and Orchestration.}
Classic IR has explored post-hoc diversification via reranking (e.g., MMR~\cite{carbonell1998use}). \alphapart{} can be seen as the systems-level dual: we enforce diversity at the candidate generation phase by partitioning computational work, rather than at the reranking phase by analyzing semantic content. Similarly, modern retrieval orchestrators like ACORN~\cite{kim2025acorn} manage multiple retrievers but do not solve inter-lane duplication within a single retriever. Our method provides the missing primitive for efficient, coordination-free parallel execution within any given component of such a system.

\section{Discussion and Limitations}

\subsection{Deployment Guide}
\label{sec:deployment}

Step 1: Measure $\rhozero$ on sample queries (100-1000 offline or via instrumented serving). Compute Jaccard overlap: $\rhozero = |\cap \text{lanes}| / |\cup \text{lanes}|$.

Step 2: Set $\alpha$ based on $\rhozero$:
\begin{itemize}
\item $\rhozero \ge 0.9$ (HNSW-like) $\to$ set $\alpha = 1.0$ $\to$ expect 2-4$\times$ quality lift
\item $0.6 \le \rhozero < 0.9$ $\to$ set $\alpha = 0.7$ $\to$ moderate gains
\item $\rhozero < 0.6$ $\to$ set $\alpha = 0.5$ $\to$ small but monotone improvement
\end{itemize}

A good default in high-overlap regimes ($\rhozero \gtrsim 0.9$) is $\alpha=1$ with $\Kpool=\kTotal$; this uses the budget the way consistent hashing would distribute keys~\cite{karger1997consistent} while preserving hedging's latency benefits~\cite{dean2013tail,primorac2021hedge}.

Step 3: Size $\Kpool = \kTotal$ (exceptions rare).

Step 4: Validate parity (log \efsearch/\nprobe, node visits).

Step 5: Monitor $\rhozero$ over time; retune if drift.

\subsection{Deployment Patterns}

\begin{itemize}
\item Hedging with extended timeout (turn late arrivals into coverage gains)
\item Intra-node parallel HNSW traversals (especially on multi-core)
\item Multi-region active-active (geographic lanes explore disjoint spaces)
\item Sharded indexes (shard-local partitioning if queries fanout)
\end{itemize}

\subsection{Stragglers and Timeouts}
\label{sec:stragglers}

\alphapart{} is coordination-free at runtime: lanes share only the PRF definition and query seed. 
If a lane times out, the union of lanes that \emph{did} return remains complementary because dedicated slices are disjoint. 
Two practical policies are common: 
(i) \emph{first-$K$ arrivals}: return as soon as merged results reach the target $k$; 
(ii) \emph{time-boxed backfill}: wait to the deadline, then backfill any shortfall from the shared suffix when $\alpha{<}1$. 
Either policy preserves hedging's latency benefits while turning late arrivals into additional coverage rather than duplicates.

\subsection{Heterogeneous Lanes}
\label{sec:heterogeneous}

When lanes have unequal budgets $k_{\text{lane},r}$ with $\sum_{r=1}^{M} k_{\text{lane},r} = \kTotal$, set 
$k_{\text{ded},r} = \lfloor \alpha \, k_{\text{lane},r} \rfloor$.
Reserve a dedicated block of size $k_{\text{ded},r}$ for each lane within the first 
$\sum_{i=1}^{M} k_{\text{ded},i}$ positions of the PRF-ordered pool so that each block occupies a distinct congruence class modulo $M$. 
By construction, dedicated selections are disjoint across lanes; the shared suffix remains a single contiguous range used for backfill when $\alpha{<}1$. The construction in Section~\ref{sec:method} extends naturally to this case.

\subsection{Limitations and Future Work}
\label{sec:limitations}

\paragraph{Algorithmic Generality and the Candidate Pool.}
The primary requirement of our method is the ability to efficiently enumerate a deterministic candidate pool larger than any single lane's budget. This makes it a natural fit for HNSW and IVF. However, this highlights a systemic constraint: algorithms that are inherently single-pass and cannot decouple candidate generation from final scoring would require fundamentally different coordination strategies. This points to a desirable property for future ANN algorithms intended for parallel deployment: the ability to cheaply generate supervisets of candidates for subsequent partitioning and rescoring.

\paragraph{Architectural Implications of Two-Phase Search.}
For HNSW, our "pool-then-partition" approach imposes a two-phase execution model. While our experiments validate cost parity in terms of node visits, a naive implementation could introduce a synchronization barrier between the phases. A key area for future engineering is to create tightly integrated, pipelined implementations that stream candidates from the pooling phase directly into the partitioning and rescoring logic, mitigating any potential latency overhead and reducing peak memory usage, especially for very large values of $\kTotal$.

\paragraph{Future Directions.}
Our work establishes computational diversity as a core design principle. The immediate next steps involve extending this principle dynamically. An online system could track $\rhozero$ in real-time and adapt the $\alpha$ parameter on a per-query or system-wide basis to respond to shifts in data or query distributions. Extending the evaluation to other convergent index families, such as DiskANN and graph quantization methods, is also a critical next step. Finally, while we emulate common production patterns, validating our measured $\rho_0$ values against live production telemetry would provide the ultimate confirmation of the problem's severity and the solution's impact.

\section{Conclusion}

Parallel lanes waste budget due to convergent traversal (quantified by $\rhozero$). \alphapart{} makes their work complementary with a zero-overlap guarantee at equal cost. The method matches single-index quality on HNSW and recovers free quality on IVF, with a simple deployment rule: measure $\rhozero$, set $\alpha$, use $\Kpool{=}\kTotal$. This establishes computational diversity as a first-class design goal for parallel ANN search.

\paragraph{Supplementary material.}
An appendix with abbreviations and implementation details is provided as a separate PDF uploaded with this submission.

\balance
\setlength{\bibsep}{2pt plus 0.3ex}
\bibliography{references}
\bibliographystyle{mlsys2025}

\end{document}